\documentclass[12pt]{article}
\usepackage{multirow}
\usepackage{multicol}
\usepackage[dvipsnames,svgnames,table]{xcolor}
\usepackage{graphicx}
\usepackage{epstopdf}
\usepackage{amsmath,amssymb,amsfonts,amsthm}
\usepackage{ulem}
\usepackage{hyperref}
\usepackage{mathrsfs}
\usepackage{physics}
\usepackage{braket}
\RequirePackage{calc}
\RequirePackage{indentfirst}
\RequirePackage{fancyhdr}
\RequirePackage{lastpage}
\RequirePackage{ifthen}
\RequirePackage{lineno}
\RequirePackage{float}
\RequirePackage{setspace}
\RequirePackage{enumitem}
\RequirePackage{mathpazo}
\usepackage{upgreek}
\author{Yuri Kamyshkov}
\title{}
\usepackage[paperwidth=612pt,paperheight=792pt,top=72pt,right=72pt,bottom=72pt,left=72pt]{geometry}

\makeatletter
{\par\setlength{\parindent
	}{#3}
	\setlength{\leftmargin}{#1}       \setlength{\rightmargin}{#2}%
	\advance\linewidth -\leftmargin       \advance\linewidth -\rightmargin%
	\advance\@totalleftmargin\leftmargin  \@setpar{{\@@par}}%
	\parshape 1\@totalleftmargin \linewidth\ignorespaces}{\par}%
\makeatother 

\def\red[#1]{{\color{red} #1}}
\def\blue[#1]{{\color{blue} #1}}

\begin{document}

\begin{center}
	\textbf{{\large On the Neutron Transition Magnetic Moment}}
\end{center}

\begin{center}
	\textrm{Zurab Berezhiani $^{1,2}$, Riccardo Biondi $^{1,2}$, Yuri Kamyshkov $^{3}$ and Louis Varriano $^{3,4}$}
\end{center}

\begin{center}
	\textit{{\small \quad $^1$ Dipartimento di Fisica e Chimica, Universit\`a di L'Aquila, 67010 Coppito AQ, Italy \\
			              \quad $^2$ INFN, Laboratori Nazionali del Gran Sasso, 67010 Assergi AQ, Italy\\
			              \quad $^3$ Department of Physics, University of Tennessee, Knoxville, TN 37996-1200, USA\\
			              \quad $^4$ Department of Physics, University of Chicago, Chicago, IL  60637, USA} }
\end{center}

\date{\today}

\begin{abstract}
	\small
	
	We discuss the possibility of the  transition magnetic moments (TMM) between the neutron $n$ and 
	its hypothetical sterile twin ``mirror neutron'' $n'$ from a parallel particle ``mirror'' sector. 
	The~neutron can be spontaneously converted into mirror neutron via the TMM 
	(in addition to the more conventional transformation channel due to $n-n'$ mass mixing)
	interacting with the magnetic field ${\bf B}$ as well as with mirror magnetic field ${\bf B}'$. 
	We derive analytic formulae for the average probability of $n-n'$ conversion and 
	consider possible experimental manifestations of neutron TMM effects.  
	In particular,  we discuss the potential role of these effects in the neutron lifetime measurement 
	experiments leading to new, testable predictions.
\end{abstract}

\section{Introduction}
In Refs.~\cite{Berezhiani:2005hv,Berezhiani:2008bc} the idea was conjectured that 
the neutron $n$ can be transformed into a sterile neutron $n'$ that belongs 
to a hypothetical parallel mirror sector.  
This mirror sector is an exact copy of the ordinary particle sector with identical fermion content and 
identical gauge forces, different in the respect  that the strong and electroweak forces described 
by the Standard Model (SM) $SU(3)\times SU(2) \times U(1)$ act only between ordinary particles, 
and gauge forces of the mirror Standard  Model (SM$'$) $SU(3)'\times SU(2)' \times U(1)'$ 
act  only between mirror particles (for a review, see e.g. Refs.~\cite{Alice}).  
The particle physics of the two sectors are exactly the  same due to a discrete symmetry $PZ_2$ 
under exchange of all ordinary and mirror particles modulo the fermion chirality which symmetry can be 
considered as a generalization of parity, namely for our sector being left-handed, the~mirror 
sector can be considered  as right-handed.  
In fact, parity restoration was the initial motivation for introducing 
the mirror sector~\cite{Mirror}. 
However, in~principle, this discrete symmetry can be simply $Z_2$ without the chirality change,  
in which case the  parallel sector will have identical physics in left basis~~\cite{Alice}).  
Most physical applications do not depend on the type of this discrete symmetry, 
and we shall continue to call this parallel sector the ``mirror sector'' in both cases. 
If   $Z_2$ or $PZ_2$ is an exact symmetry, 
then each ordinary particle as the electron, photon, proton, neutron etc. must have a 
mirror twin, the~electron$'$, photon$'$, proton$'$, neutron$'$ etc. exactly degenerate in mass.  
Interactions between the two sectors are possible  via the common gravitational force
but, in~principle, also via some very feeble interactions induced by new physics beyond the Standard Model.  
These new interactions, typically related to higher order effective operators, 
can arise at some a priori unknown energy scale which in principle might be not very far
and can be even as small as few TeV.   
Any new interactions must respect gauge invariances of both sectors and can manifest itself 
in a mixing phenomena between the neutral particles of the two sectors, 
in particular as  the photon-mirror photon kinetic mixing~\cite{Holdom} 
the mixing of (active)  ordinary neutrinos and (sterile)  mirror neutrinos~\cite{ABS}, 
the~above mentioned mixing of the neutron and mirror neutron~\cite{Berezhiani:2005hv,Berezhiani:2008bc},  
and similar mixings between other neutral particles such as pions and kaons induced 
e.g., by the common gauge flavor symmetry between the two sectors~\cite{MFV}. 
Mirror matter is a viable candidate for light dark matter, a~sort of asymmetric atomic dark matter 
consisting dominantly of mirror hydrogen and helium~\cite{BCV}. 
Therefore, any transition of a neutral particle of the ordinary sector into a neutral particle 
of another sector can be considered as  a conversion to dark matter. 
In fact, the~underlying baryon or lepton number (and CP) violating particle processes in the early universe
can be at the origin of the cogenesis of ordinary and dark baryon asymmetries~\cite{BB-PRL}. 
However, in~this paper we will not focus on the cosmological aspects of mirror matter  
(see Refs.~\cite{BCV,BCV1,AntiDM}  
and reviews ~~\cite{Alice} 
for corresponding discussions) 
and shall discuss essentially  the features of a laboratory search  for $n \rightarrow n'$ transition. 

This transition was suggested to occur via the mass mixing term 
$\epsilon_{nn'} \overline{n}n' + {\rm h.c.}$ in Ref.~\cite{Berezhiani:2005hv},   
and~the masses of $n$ and $n'$ are exactly 
the same due to the initial assumption of mirror parity leading to the 
the degeneracy between the ordinary and mirror particles. 
Then, the~phenomenon of free $n-n'$ oscillation is essentially described  in the same way 
as free neutron--antineutron oscillation $n-\bar n$ due to the Majorana mass term 
$\epsilon_{n\bar n} n^T C n + {\rm h.c.}$ \cite{Kuzmin}~(for  a recent reviews 
see~\cite{Phillips}). 
Namely, for~free non-relativistic neutrons in vacuum but in the presence of magnetic fields, 
the time evolution of $n-\bar n$ and $n-n'$ systems are described respectively 
by the Hamiltonians\footnote{Hereafter we use natural units, $\hbar =1$, $c=1$. } 
\begin{equation}\label{eq:1}
{\hat H}_{n\bar n}  =
\begin{pmatrix}
m + \mu  \boldsymbol{\sigma} \bf{B} & \epsilon_{n\bar n} \\
\epsilon_{n\bar n}  & m - \mu  \boldsymbol{\sigma} \bf{B}\\
\end{pmatrix},  \quad\quad \quad 
{\hat H}_{nn'}  =
\begin{pmatrix}
m + \mu \boldsymbol{\sigma} \bf{B} & \epsilon_{nn'}\\
\epsilon_{nn'} & m'  + \mu'  \boldsymbol{\sigma} \bf{B}' \\
\end{pmatrix}.
\end{equation}
where $m,m' $ and $\mu, \mu'$ are respectively the masses and magnetic 
moments of the neutron and mirror neutron, ${\bf B}$ 
and  ${\bf B}'$ are ordinary and mirror magnetic fields, and~ 
$\boldsymbol{\sigma} = (\sigma_1,\sigma_2,\sigma_3)$ stands for Pauli~matrices. 

However,  between~these two cases there are important differences: 
\begin{itemize}[leftmargin=10mm,labelsep=6mm]
\item[(i)] First of all, $n-\bar n$ transition changes the baryon number $B$ by two units, $\Delta B =-2$, 
while $n-n'$ transition changes $B$ by one unit, $\Delta B=-1$, but~it changes 
also the mirror baryon number $B'$ by one unit, $\Delta B' = +1$. 
Therefore, $n-n'$ mixing conserves the combination of two baryon numbers $\bar B= B+B'$  
while  $n-\bar n$ mixing violates $\bar B$.   
From the theoretical side, the phenomena of neutron--mirror neutron and neutron--antineutron 
mixings can be intimately related, with $\bar B$--conserving  $n-n'$ mixing  being a dominant 
effect and $n-\bar n$ mixing being subdominant effect emerging due to explicit \cite{Berezhiani:2005hv} 
or spontaneous \cite{BM} violation of $\bar B$. 

\item[(ii)] Exact degeneracy between the neutron and antineutron masses (and also magnetic moments) 
is based on fundamental CPT invariance, which cannot be violated in the frames of 
local relativistic field theories.  (However, environmental  energy 
splitting can be induced by some long range fifth-forces 
related e.g.,~to very light $B-L$ baryophotons~\cite{BL}, or~in bi-gravity picture when 
ordinary and mirror components are  coupled to different metric tensors~\cite{bigrav}.) 
The degeneracy between the neutron and mirror neutron is related to mirror 
parity which in principle can be spontaneously broken~\cite{BDM}. 
The order parameters of this breaking can be naturally small and the mass splitting 
between $n$ and $n'$ states 
can be rather tiny, say as small as $10-100$~neV  in which case it can have 
implications for the neutron lifetime problem~\cite{n-life}. 
Both $n-\bar n$ and $n-n'$ oscillations are affected by the matter medium and magnetic 
fields. However, in~the case of $n-n'$ oscillation, the~presence of mirror matter and 
mirror magnetic field ${\bf B}'$ will be manifested as uncontrollable background in~experiments. 

\item[(iii)] $n-\bar n$ transition can be experimentally manifested~\cite{Phillips} 
as the antineutron appearance in the 
beam of free neutrons, or~as nuclear disintegration $(A,Z) \to (A-2, Z-\Delta Z) + \pi$'s  
due to $n \to \bar n$ conversion of a neutron bound in nuclei and its subsequent annihilation
with other nucleons producing multiple pions. 
As for $n \to n'$ transition, it is kinematically suppressed  for a bound neutron, 
simply because of energy conservation, and~thus it has no influence on the stability 
of nuclei~\cite{Berezhiani:2005hv}. 
However,  free $n-n'$ transition is possible and it  can be experimentally manifested 
as the neutron disappearance $n\to n'$ or regeneration $n\to n' \to n$ \cite{Berezhiani:2005hv}.

\item[(iv)] There are severe experimental limits on $n-\bar n$ mixing mass $\epsilon_{n\bar n}$, 
usually expressed as limits on the free oscillation time 
$\tau_{n\bar n} = 1/\epsilon_{n\bar n}$. 
Namely, the~direct experimental limit on free $n\rightarrow \bar{n}$ oscillation is
$\tau_{n\bar n} > 0.86 \times 10^8$~s~\cite{BaldoCeolin:1994jz}
while the limit from the nuclear stability~\cite{Abe:2011ky} yields $\tau_{n\bar n} > 2.7 \times 10^8$~s. 
The latter corresponds to the upper bound $\epsilon_{n\bar n} < 2.5\times 10^{-24}$~eV.  
As for $n-n'$ oscillation, it can be rather fast, even faster than the neutron decay itself~\cite{Berezhiani:2005hv}.
Several dedicated experiments were performed for testing the ultra-cold neutron (UCN) 
disappearance due to $n\to n'$ transition~\cite{Ban,Serebrov1,Serebrov2,Bodek,Altarev,ILL}.
Assuming~the absence of the mirror magnetic field at the Earth, the~strongest limit was 
obtained in Ref.~\cite{Serebrov2} that implies $\tau_{nn'} > 448$~s which is equivalent 
to an upper limit $\epsilon_{nn'} < 1.5 \times 10^{-18}$~eV.  
This limit, however, becomes invalid if the Earth possesses a mirror magnetic field $B'$~\cite{Berezhiani:2008bc}. 
For non-zero $B'$ the limits on $\tau_{nn'}$  are much weaker, 
and~the present experimental  situation is summarized in Ref.~\cite{ILL}. 
In~fact, for~$B'>0.3$~G, $n-n'$ oscillation time as small as $\tau_{nn'} \sim 1$~s can be allowed. 
In addition,  some of the experimental data show  significant anomalies,  
the~strongest  one  of $5.2\sigma$ deviation from the null hypothesis~\cite{Nesti}, 
which can be interpreted by $n-n'$ oscillation with $\tau_{nn'} \sim 10$~s 
(or $\epsilon_{nn'} \sim 10^{-16}$~eV) 
in the presence of mirror magnetic field $B' \sim 0.1$~G.  
Mirror magnetic field at the Earth could be induced by a tiny fraction of captured mirror matter 
via the electron drag mechanism~\cite{BDT}. 
Let us also remark that fast $n-n'$ oscillation can have interesting implications 
for the propagation of ultra-high energy cosmic rays at  cosmological distances~\cite{Gaz} 
or for the neutrons from solar flares \cite{flares}.  
These oscillations also can be tested via $n\to n'\to n$ regeneration experiments like 
those discussed in Refs.~\cite{Berezhiani:2017azg,Broussard:2017yev} and 
will be discussed also in this~paper in relation to the possible existence of transition magnetic moment 
between $n$ and $n'$ states.   

\end{itemize} 

The neutron magnetic dipole moment $\mu$ determines 
the Larmor precession of the neutron spin in an external magnetic field $\bf B$. 
It is known with high precision, $\mu = (-1.91304273 \pm 0.00000045) \, \mu_N$ \cite{PDG}
where $\mu_N = e/2m_p$ is the nuclear magneton. 
It is convenient to use it as $\mu =  -6 \times 10^{-12}$ eV/G. 
In principle, the~neutron could have also an electric dipole moment which  
would violate P and CP-invariance; however, there are severe experimental limits on it~\cite{PDG}. 
If the mirror sector exists, the~same limits on an electric dipole moment should apply to the mirror~neutron.

However, the~transformation $n-n'$ can be  also due the existence of a transition 
magnetic moment (TMM) between the neutron and mirror neutron.
Notice that the existence of a TMM between the neutron and antineutron 
is forbidden by Lorentz invariance while  between 
the neutron and mirror neutron it is allowed~\cite{Arkady}. 
In other terms, there can exist the following non-diagonal operators between $n$ and $n'$ states
\begin{equation} \label{eta-tr} 
\eta \, F_{\mu\nu} \, \overline{n} \sigma^{\mu\nu} n'  + 
\eta'  \, F'_{\mu\nu} \, \overline{n} \sigma^{\mu\nu} n' ~ + ~ {\rm h.c.}
\end{equation} 
where $F_{\mu\nu}$ and $F'_{\mu\nu}$ are respectively the ordinary and mirror 
electromagnetic fields. For~the transition magnetic moments (TMM) we have 
$\eta' = \pm \eta$ depending on the type of exchange symmetry between 
two sectors, $Z_2$ or $PZ_2$. Let us notice that the constants $\eta$ in Eq.~(\ref{eta-tr}) 
can be generically complex, and besides the TMM there can exist also transition electric dipole moment 
(TEDM)  between $n$ and $n'$ states which potentially would introduce CP violating effects 
in $n-n'$ conversion. 

In this paper we are interested in exploring possible experimental manifestations of the neutron TMM.
We do not discuss the particular mechanisms of its generation. 
Most generically, if~a mechanism exists that causes the $n-n'$ mixing, 
it will involve the charges of the constituent quarks of $n$ and $n'$ and the corresponding  
interactions of these charges either with  the photons or with mirror photon will induce also 
the transitional moments between $n$ and $n'$.  
Moreover, the~neutron TMM can be induced  by loops involving hypothetical charged particles, 
in an analogous way as the transition magnetic moment between neutrinos 
(for~some possible models  one can address Ref.~\cite{LHEP}).  
In principle, such loops should induce both the TMM and TEDM with the comparable magnitudes. 
However, here for brevity we concentrate on the case of transition magnetic moment~only. 


\section{Neutron TMM and \boldmath{$n-n'$} System} \label{sec2}

For describing the time evolution of the mixed $(n, n')$ system 
in the background of uniform magnetic fields ${\bf B}$ and ${\bf B}'$ and the
possible presence of ordinary and/or mirror matter, 
the Hamiltonian $H_{nn'}$ of Equation~(\ref{eq:1}) should be modified to the following form 
\begin{equation}\label{H1}
{\hat H}_{nn'}  =
\begin{pmatrix}
m + V + \mu  {\bf{B} } {\boldsymbol{\sigma} } & 
\varepsilon+ \eta \mathbf{B}  {\boldsymbol{\sigma} } + \eta'  \mathbf{B}'  \boldsymbol{\sigma} \\
\varepsilon+\eta \mathbf{B} \boldsymbol{\sigma} + \eta'  \mathbf{B}'  \boldsymbol{\sigma} & 
m' + V' + \mu'  \bf{B'} \boldsymbol{\sigma} \\
\end{pmatrix}  \, , 
\end{equation}
where we neglected the terms describing the decay, incoherent scattering and absorption of $n$ and $n'$ states
	assuming that the densities of both ordinary and mirror matter  are low. 
	However we kept the optical potentials due to coherent zero-angle scattering.  

Hamiltonian (\ref{H1}) is in fact a $4\times 4$ 
Hermitian matrix that includes two spin-polarizations of $n$ and $n'$ states. 
Here $V$ and $V'$ stand for $n$ and $n'$ Fermi potentials induced respectively 
by ordinary and mirror matter, $\boldsymbol{\sigma}$ is a set of Pauli matrices, 
$\varepsilon = \epsilon_{nn'}$ is a mass mixing of Equation~(\ref{eq:1}) and 
$\eta$ and $\eta'$, 
are the TMM's between $n$ and $n'$ of Equation~(\ref{eta-tr}) 
related respectively to ordinary and mirror magnetic fields.  
In the following, in~view of $Z_2$ or $PZ_2$ parities, 
we take $\mu' = \mu =  -6 \times 10^{-12}$ eV/G  for normal magnetic moments of $n$ and $n'$ 
while for the TMM's we consider two possibilities, $\eta' = \pm \eta$.  

The magnitude of $n$TMM $\eta$ is unknown; it can be either positive or negative. 
(Let us remind that for simplicity we do not discuss transitional electric dipole moments.)    
We can measure the TMM in units of the neutron magnetic moment itself, $\eta = \kappa \mu$  
with $\kappa = \eta/\mu \ll 1$ being dimensionless parameter. 
Therefore, Equation~(\ref{H1}) can be conveniently rewritten as
\begin{equation}\label{H2}
{\hat H}_{nn'} =
\begin{pmatrix}
2 \boldsymbol{\omega} \boldsymbol{\sigma}  + 2\delta 
& \varepsilon + 2\kappa (\boldsymbol{\omega} \pm \boldsymbol{\omega}'  )\boldsymbol{\sigma} \\
\varepsilon + 2\kappa (\boldsymbol{\omega} \pm \boldsymbol{\omega}' \boldsymbol{\sigma}  )& 
2\boldsymbol{\omega}'  \boldsymbol{\sigma} \\
\end{pmatrix} 
\end{equation}
where 
$2\boldsymbol{\omega} = \mu {\bf B}$,  $2\boldsymbol{\omega}' = \mu {\bf B}'$ 
and $2\delta = (V-V') + (m-m')$  
(one can drop the equal additive diagonal terms since they will not affect the evolution of the system), 
and the $\delta$--term  which can be positive or negative 
comprises the difference of  the Fermi potentials  as well as the possible  mass difference 
between the ordinary and mirror neutrons  as discussed in Ref.~\cite{n-life}. 
The sign $\pm$ in the non-diagonal terms takes into account that the unknown parity 
of  the mirror photon ($B'$) can be the same (+) or opposite ($-$) to that of the ordinary photon. 
For simplicity, we coin these two cases as $+$ and $-$ parities.  

Let us consider first the case when mirror magnetic field is negligibly small, $B' \approx 0$ 
(e.g.,~mirror~galactic magnetic fields of few $\upmu$G can be the case here)  but ordinary 
magnetic field is rather large and it cannot be neglected.  In~ $B' = 0$ approximation 
(or $B' \ll  B$) the time evolution described by the Hamiltonian (\ref{H2}) is very simple since 
the spin quantization axis can be taken in the direction 
of magnetic field ${\mathbf B} = (0,0,B)$ thus reducing the evolution Hamiltonian  
($4\times 4$ matrix)  to two independent $2\times 2$ matrices   for two polarization states:
\begin{equation}\label{22}
{H}^\uparrow_{nn'} =
\begin{pmatrix}
2(\delta - \omega) & \varepsilon - 2\kappa\omega \\
\varepsilon - 2\kappa\omega & 0  \\
\end{pmatrix}, \quad\quad 
{H}^\downarrow_{nn'} = \begin{pmatrix}
2(\delta+ \omega) & \varepsilon + 2\kappa\omega \\
\varepsilon + 2\kappa\omega & 0  \\
\end{pmatrix}, 
\end{equation} 
where signs $^\uparrow$ and $^\downarrow$  stand for the cases 
of the neutron spin parallel/antiparallel  
to the magnetic field direction ${\mathbf B}$, and~since $\mu =  -6 \times 10^{-12}$ eV/G 
we have numerically $2\omega = \vert \mu B \vert = 9\times(B/1~{\rm mG})~{\rm s}^{-1}$. 
Therefore, generically the probabilities of 
$n \to n'$  transition  after time $t$  depend on the neutron~spin-polarizations:
\begin{eqnarray}\label{delta}
&& P^{\uparrow}(t) = 
\frac{(\varepsilon - 2\kappa \omega)^2}{(\delta - \omega)^2 + (\varepsilon - 2\kappa \omega)^2} 
\sin^2[ \sqrt{ (\delta -\omega)^2 + (\varepsilon - 2\kappa \omega)^2}~t] \, , \nonumber \\ 
&& P^\downarrow(t) =
\frac{(\varepsilon + 2\kappa \omega)^2}{(\delta + \omega)^2 + (\varepsilon + 2\kappa \omega)^2} 
\sin^2[\sqrt{(\delta + \omega)^2 + (\varepsilon + 2\kappa \omega)^2}~t]\, .      
\end{eqnarray} 

In particular, if~$\delta\neq 0$, one can tune the applied magnetic field $B$ 
(i.e., $2\omega = \vert \mu B \vert $) so that $\omega \approx \vert \delta \vert$. 
Then the $n-n'$ conversion probability  can be resonantly enhanced 
for one polarization state ($+$ or $-$, depending on the sign of $\delta$ 
and also  on $\kappa$ once the $n$TMM effects are included). 

Let us consider the case when $\delta=0$ which corresponds to the minimal hypothesis 
assuming that $n$ and $n'$ are exactly degenerate in mass, there are no effects of mirror matter 
at the Earth, and~the ordinary gas density is properly suppressed as is usually done 
in the experiments using the UCN for the neutron lifetime measurements of for search 
of $n-n'$ effects.  Then oscillation probabilities~(\ref{delta})~become
\begin{equation}\label{updown}
P^{\uparrow}(t) = \frac{(\varepsilon - 2\kappa \omega)^2}{\omega^2} \sin^2(\omega t), \quad \quad 
P^\downarrow(t) = \frac{(\varepsilon + 2\kappa \omega)^2}{\omega^2} \sin^2(\omega t),     
\end{equation} 
where we consider that $\varepsilon \ll \omega$ and $\kappa \ll 1$.  
In the general case, when both $\varepsilon$ and $\eta$ are present, 
these probabilities are different.  
However, if~$\eta =0$ (or $\varepsilon=0$), $P^{\uparrow}(t)$ and $P^{\downarrow}(t)$ must be equal. 

For~unpolarized neutrons one can consider an average probability between two polarisations:
\begin{equation}\label{time}
P(t) = \frac12 \big[P^{\uparrow}(t) + P^\downarrow(t) \big] = 
\frac{\varepsilon^2}{\omega^2} \sin^2(\omega t) \, + \, 4\kappa^2 \sin^2(\omega t) 
= P_\varepsilon(t) +  P_\eta(t) 
\end{equation} 
In this case contributions of $n-n'$ mass mixing and magnetic moment mixing are disentangled, 
and they  contribute independently as $P_\varepsilon(t)$ and $P_\eta(t)$. 
Obviously, this is not the case when $\delta\neq 0$ as one can  see directly from Equation~(\ref{delta}).   

The typical time for free neutron propagation in experiments is $t < 0.1$~s or so 
(e.g., for ultra-cold neutrons (UCN) this is the time between bounces from the walls in the trap). 
Thus, provided that $\omega t \gg 1$, which means that $B$ is larger than few mG  
(which regime we call the case of non-zero field $B\neq 0$), 
the oscillation probabilities can be averaged in time  
and  we obtain
\begin{equation}\label{Bnon0}
\overline{P}_\varepsilon(B\!\neq \!0,B'\!=\!0)  = \frac{\varepsilon^2}{2\omega^2}\, , \quad \quad 
\overline{P}_\eta(B\!\neq\! 0,B'\!=\!0) = 2\kappa^2 .   
\end{equation} 

However, in~small magnetic field $B < 1$~mG or so, when $\omega t \ll 1$ 
(which case we can call zero-field limit),  oscillations cannot be averaged, and~we have
\begin{equation}\label{B0}
P_\varepsilon(B\!\to\! 0,B'\!=\!0)  =  \varepsilon^2 \langle t^2 \rangle ,   \quad \quad 
P_\eta(B\!\to\! 0,B'\!=\!0) = 4\kappa^2 \omega^2  \langle t^2 \rangle,  
\end{equation} 
where $\langle t^2 \rangle$ stands for the mean free-flight time square 
averaged over the neutron velocity spectrum. 
Therefore, the~two cases can be distinguished by comparing the neutron losses 
in zero and non-zero field regimes. In~the case of $n-n'$ mass mixing $\varepsilon$, the~non-zero field 
suppresses $n \to n'$ transitions, 
$\overline{P}_\varepsilon(B\!\neq \!0) < P_\varepsilon(B\!\to\! 0)$, and~
thus the neutron losses should be larger in smaller applied field. 
In the case of $n-n'$ TMM, the~situation is just the opposite, 
$\overline{P}_\eta(B\!\neq \!0) > P_\eta(B\!\to\! 0)$, 
so that in the limit of zero magnetic field  $n-n'$ conversion probability due to $n$TMM is suppressed 
while for large enough $B$ it  becomes constant. 
Summarizing, in~the generic case, both effects can contribute in $n-n'$ transitions 
and  the average oscillation probability has a form 
$ \overline{P} =  \overline{P}_\varepsilon  + \overline{P}_\eta =  (\varepsilon/\sqrt2\omega)^2+ 2\kappa^2$. 
Thus, if there is no mirror magnetic field at the Earth,  $B'=0$,
for smaller applied field $B$, the~contribution of the first term should be larger, whereas for 
large enough magnetic field, when $2\kappa \omega > \varepsilon$, the~second term 
induced by the~$n$TMM effect should become dominant.  

However, in~the presence of non-zero mirror magnetic field, $B'\neq 0$, the~situation becomes different. 
Let us consider the case when the mirror magnetic field  at the Earth is not negligible, 
i.e., it is at least larger than few mG and it perhaps can be $\sim 1$~G, 
comparable with the normal magnetic field of the Earth~\cite{Berezhiani:2008bc}.  
In the presence of $n-n'$ mass mixing but without the $n$TMM terms,  the~exact general solution 
for $n-n'$ oscillation probability vs. time  in a constant and uniform magnetic fields 
${\mathbf B}$ and ${\mathbf B}'$ was obtained in Refs.~\cite{Berezhiani:2008bc,Nesti}.
The probability of $n-n'$ conversion 
in this case has a resonance character at $|\bf{B}|=|\bf{B'}|$ and it depends on the spatial angle 
between vectors $\bf{B}$ and $\bf{B'}$. The~direction and magnitude of the vector $\mathbf{B'}$ 
is a priori unknown but it can be determined from  scanning experiments, 
e.g.,~with the high-flux cold neutron beams as described in
Refs.~\cite{Berezhiani:2017azg,Broussard:2017yev}. 


Let us now study a situation with $B'\neq 0$ 
when also the $n$TMM term is present in the Hamiltonian~(\ref{H2}) 
(but assuming again  that $\delta=0$).  
Now the probability of $n-n'$ transition depends on the values of $\varepsilon$ as well as $\kappa$. 
The exact expression for time dependent probability is difficult to extract in analytical form from
the Hamiltonian represented by $4\times 4$ matrix (\ref{H2}).  
However, if~we focus on non-resonant region assuming that $\vert \omega-\omega'\vert t \gg 1$, 
which for the neutron mean flight times $t \sim 0.1$ s 
means that the difference of magnetic fields $\vert B-B' \vert $ is  larger than several mG, 
then the~mean oscillation probability 
$\overline{P}= \frac12 (\overline{P}^\uparrow + \overline{P}^\downarrow)$  between 
two spin states, averaged over many oscillations, can~be readily calculated 
following the techniques of Ref.~\cite{Berezhiani:2008bc}. 
Interestingly,  the~interference terms between the
two effects cancel out and one gets the average probability simply as a sum 
$\overline{P} = \overline{P}_\varepsilon + \overline{P}_\eta$. 

Here $\overline{P}_\varepsilon$ is the average $n-n'$ oscillation probability due to mass mixing: 
\begin{equation}\label{Peps}
\overline{P}_\varepsilon({\mathbf B}) 
= \frac{\varepsilon^2   }{2(\omega-\omega^{\prime })^2} \cos^2(\beta/2) 
+ \frac{\varepsilon^2 }{2(\omega+\omega^{\prime })^2} \sin^2(\beta/2) 
\end{equation}
where $2\omega = \vert \mu B \vert = (B/1~{G}) \times 9000~{\rm s}^{-1}$, 
analogously $2\omega' = \vert \mu B' \vert$, and $\beta$ is the angle between the 
directions of ordinary and mirror magnetic fields, ${\mathbf B}$  and ${\mathbf B}'$.   
In Refs.~\cite{ILL,Nesti} this formula was given in somewhat different equivalent form  
$\overline{P}_\varepsilon({\mathbf B}) 
= \overline{\cal P}_\varepsilon(B) + \overline{\cal D}_\varepsilon(B) \cos \beta$, 
where
\begin{equation}\label{Nesti} 
\overline{\cal P}_\varepsilon(B) =  \frac14\left[ \frac{\varepsilon^2  }{(\omega-\omega^{\prime })^2} 
+ \frac{\varepsilon^2 }{(\omega+\omega^{\prime })^2} \right]  , \quad\quad 
\overline{\cal D}_\varepsilon(B)  = \frac14\left[ \frac{\varepsilon^2  }{(\omega-\omega^{\prime })^2} 
- \frac{\varepsilon^2 }{(\omega+\omega^{\prime })^2} \right]  
\end{equation} 
which expressions depend only on the modulus of the magnetic field $B = \vert {\mathbf B} \vert$. 

The second term  $\overline{P}_\eta$  instead describes the probability of $n-n'$ 
transition only due to transition magnetic moment $\eta$,  
and it depends on the choice of $\pm$ parity. Namely, in~the case of $(-)$ parity, 
it  does not depend on the values $B$ and $B'$ and angle $\beta$, 
and~one simply gets
$\overline{P}_{\eta}^{-}({\mathbf B})=2\kappa^2$. 

As for  the case of $(+)$ parity, the~dependence on the magnetic field is non-trivial. 
Performing~calculations similar to that of Ref.~\cite{Berezhiani:2008bc},  we get for the average 
$n-n'$ transition probability
\begin{equation}\label{Ppm} 
\overline{P}_{\eta}^{+}({\mathbf B})  
= \frac{2\kappa^2 (\omega+\omega^{\prime })^2  }{(\omega-\omega^{\prime })^2} \cos^2(\beta/2) 
+ \frac{2\kappa^2 (\omega-\omega^{\prime })^2  }{(\omega+\omega^{\prime })^2} \sin^2(\beta/2) \, . 
\end{equation} 

 Obviously, these formulas for $\overline{P}_{\eta}^{\pm}$ as well as Eq.~(\ref{Peps}) 
for  $\overline{P}_{\varepsilon}$ cannot be used 
	exactly at the resonance when $\omega=\omega'$. However, in~practical sense, they are valid 
	also in proximities of the resonance as soon as $\vert \omega - \omega'\vert \gg \varepsilon$.
Let us remark also that the separability of two effects,  
$\overline{P} = \overline{P}_\varepsilon + \overline{P}_\eta$,  holds only if $\delta$ is vanishing. 
For~$\delta \neq 0$, which is the case when ordinary or mirror  matter 
densities are not negligible (or there is a mass difference between $n$ and $n'$ states)   
the effects of mass mixing and TMM cannot be separated and 
	the oscillation probability has the form  
	$\overline{P} = \overline{P}_\varepsilon + \overline{P}_\eta + \overline{P}_{\varepsilon\eta}$   
	where the ``interference'' term  
	$\overline{P}_{\varepsilon\eta}$ non-trivially depends on all parameters,   
	and generically two resonances can exist~\cite{Berezhiani:2008bc}.  

Therefore, in the case $\delta=0$, the~average probability 
in the background of mirror magnetic field ${\mathbf B}'$
can be presented as
\begin{equation}\label{PB} 
\overline{P}({\mathbf B}) = \overline{P}_\varepsilon ({\mathbf B})+ \overline{P}_\eta({\mathbf B})
= \overline{\cal P}(B) + \overline{\cal D}(B) \cos \beta \, , 
\end{equation}
where the terms $\overline{\cal P}(B)=\overline{\cal P}_\varepsilon(B)  + \overline{\cal P}_\eta(B)$ and 
$\overline{\cal D}(B)= \overline{\cal D}_\varepsilon(B)  + \overline{\cal D}_\eta(B)$, 
with $\overline{\cal P}_\varepsilon(B)$ and $\overline{\cal D}_\varepsilon(B)$ 
given in Equation~(\ref{Nesti})  and $\overline{\cal P}_\eta(B)$ and $\overline{\cal D}_\eta(B)$ 
having the following form 
\begin{equation}\label{Final} 
\overline{\cal P}_\eta^\pm (B) = 
\kappa^2 \left[ \frac{ (\omega \pm \omega^{\prime })^2 }{(\omega-\omega^{\prime })^2} + 
\frac{ (\omega \mp \omega^{\prime })^2 } {(\omega+\omega^{\prime })^2} \right], 
\quad 
\overline{\cal D}_\eta^\pm(B)  = \kappa^2
\left[ \frac{ (\omega \pm \omega^{\prime })^2 }{(\omega-\omega^{\prime })^2} - 
\frac{\ (\omega \mp \omega^{\prime })^2 } {(\omega+\omega^{\prime })^2} \right] \, . 
\end{equation} 

The values $\overline{\cal P}(B)$ and $\overline{\cal D}(B)$ can be conveniently measured 
in experiments by studying the dependence of the neutron losses on magnetic field.  
In particular, in a background of non-zero mirror field $\mathbf{B}'$  
the oscillation probabilities in the applied magnetic fields of opposite directions, 
${\mathbf B}$ and $-{\mathbf B}$,  are respectively 
$\overline{P}({\mathbf B}) = \overline{\cal P}(B) + \overline{\cal D}(B) \cos \beta$ 
and $\overline{P}(-{\mathbf B}) = \overline{\cal P}(B) + \overline{\cal D}(B) \cos (\pi-\beta)$. 
Therefore, the~average of these probabilities 
should not depend on the angle $\beta$ but only on the magnetic field value $B$. Namely, we have  
$\frac12\big[\overline{P}({\mathbf B}) + \overline{P}(-{\mathbf B})\big] =  \overline{\cal P}(B)$,  
while their difference depends also on the angle $\beta$ and we have 
$\frac12\big[\overline{P}({\mathbf B}) - \overline{P}(-{\mathbf B})\big] = \overline{\cal D}(B) \cos\beta$. 
Notice that  Equations~(\ref{Peps}) and (\ref{Final}) are invariant under 
exchange $\omega \leftrightarrow \omega'$ by mirror symmetry for both types of parity ($+$ or $-$).

In the limit $B\to 0$, when the ordinary magnetic field is screened, 
we see from Equation~(\ref{PB})   that $\overline{\cal D}(0) = 0$  
and so we get 
$\overline{P}(0) = \overline{\cal P}(0) =  \overline{\cal P}_\varepsilon (0) +  \overline{\cal P}_\eta(0)$,  
where
\begin{equation}\label{B=0} 
\overline{\cal P}_\varepsilon(0) = \overline{P}_\varepsilon(B\!=\!0,B'\!\neq\!0)
= \frac{\varepsilon^2}{2\omega^{\prime2} }, \quad  \quad
\overline{\cal P}_\eta (0)= \overline{P}_\eta (B\!=\!0,B'\!\neq\!0) = 2\kappa^2 \, .    
\end{equation} 
These expressions are analogous to that of Equation~(\ref{Bnon0}) when 
the roles of ordinary and mirror magnetic fields are reversed, i.e.,~$B \leftrightarrow B'$. 

Thus, if~$B'\neq 0$, the~probability $\overline{P}(0)$ of $n-n'$ oscillation in vanishing 
magnetic field  can be small, and~it should be modified in larger magnetic fields. 
In particular, it can be resonantly enhanced when the applied magnetic field has the values 
comparable to $B'$.  From~Equations~(\ref{B0}) and (\ref{Final}) we obtain the modification factor 
simply as a function of $y = \omega/\omega' = B/B'$:
\begin{eqnarray}\label{y} 
&& \frac{\overline{\cal P}_\varepsilon(B)}{\overline{\cal P}_\varepsilon(0)} = \frac{y^2+1}{(y^2-1)^2} \, , 
\quad \quad \quad 
\frac{\overline{\cal P}_\eta^+(B)}{\overline{\cal P}_\eta(0)} = 
\frac{y^4+6y^2 +1}{(y^2-1)^2} \, , 
\quad \quad \quad 
\frac{\overline{\cal P}_\eta^-(B)}{\overline{\cal P}_\eta(0)} = 1 \, , 
\nonumber \\
&&\frac{\overline{\cal D}_\varepsilon(B)}{\overline{\cal P}_\varepsilon(0)} = \frac{2y}{(y^2-1)^2} \, , 
\quad \quad \quad
\frac{\overline{\cal D}_\eta^+(B)}{\overline{\cal P}_\varepsilon(0)} = \frac{4(y^3+y)}{(y^2-1)^2} \, , 
\quad \quad \quad \quad 
\frac{\overline{\cal D}_\eta^-(B)}{\overline{\cal P}_\eta(0)} = 0 \, . 
\end{eqnarray} 

\begin{figure}[t]
\centering
	\includegraphics[width=7.5cm]{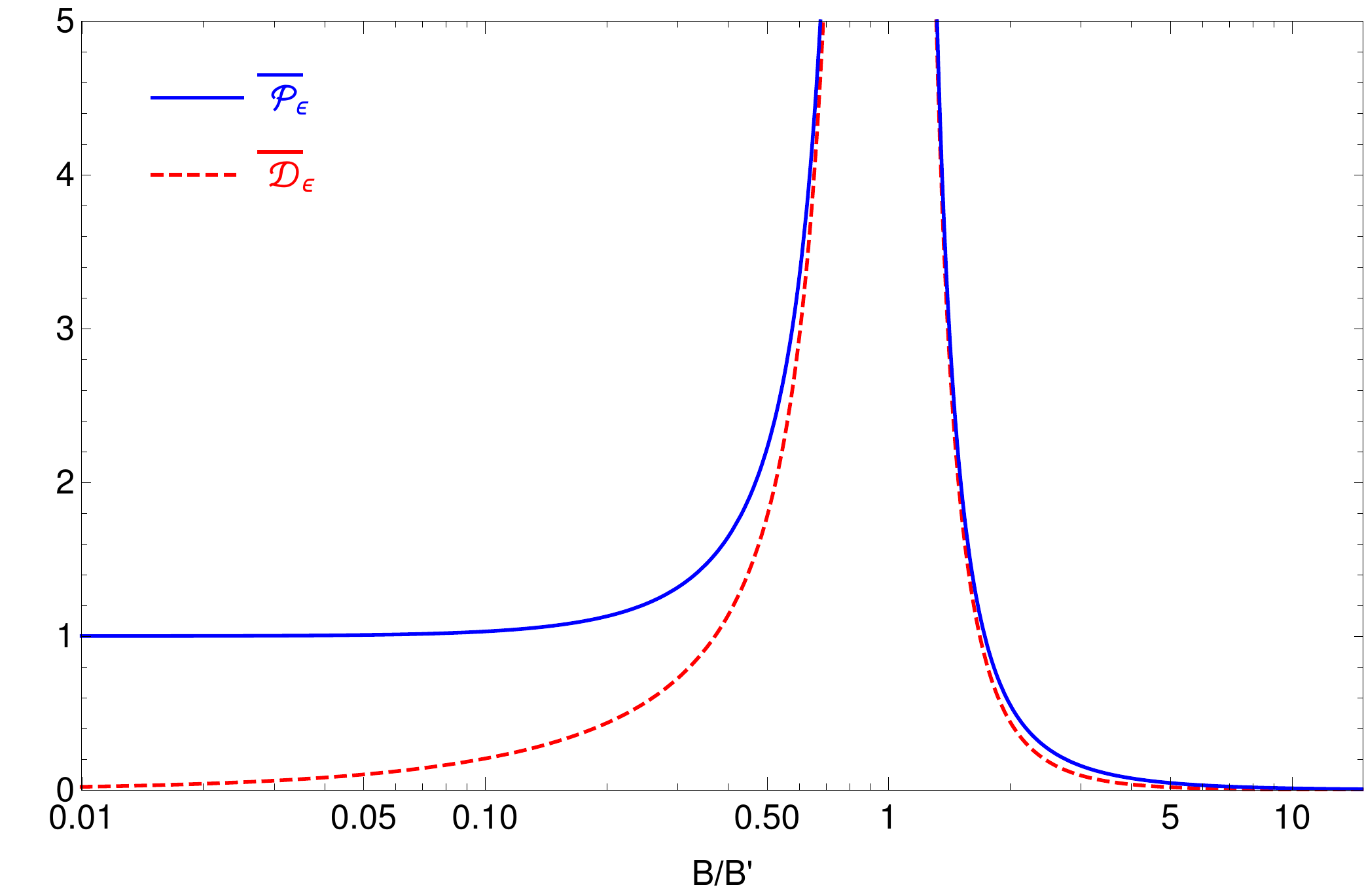}
	\includegraphics[width=7.5cm]{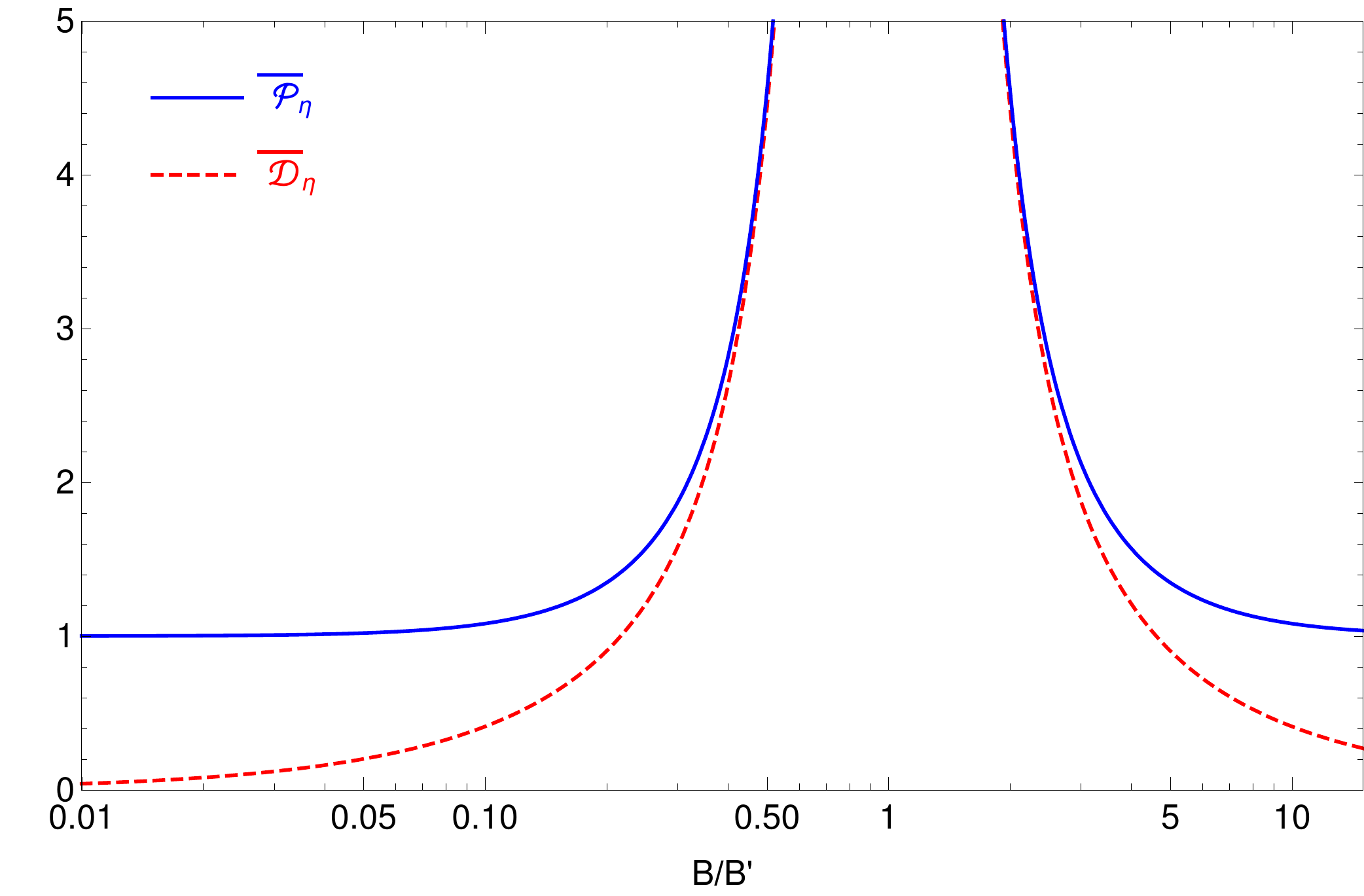}
	\caption{\label{fig:prob} 
		$\overline{\cal P}_\varepsilon(B)$ (\textbf{blue}) and 
		$\overline{\mathcal{D}}_\varepsilon(B)$ (\textbf{red}) normalized over 
		$\overline{\cal P}_{\varepsilon}(0) = \varepsilon^2 / 2\omega^{\prime2}$  {\it (\textbf{left pane})}  vs.  
		 $\overline{\cal P}_\eta^+(B)$ (\textbf{blue}) and $\overline{\mathcal{D}}_\eta^+(B)$ (\textbf{red}) 
		normalized over $\overline{\cal P}_{\eta}(0) = 2 \kappa^2$ {\it (\textbf{right panel})} 
		as functions of the ratio $y=B/B'$,   see~Equations~(\ref{y}).  
	}
\end{figure}

In principle, the~scenarios of mass mixing and $n$TTM (in the case of $+$ parity)
can be distinguished 
by the shape of the oscillation probability in the neighbourhoods of the resonance 
(see Figure~\ref{fig:prob}),  which~can be studied by scanning over the magnetic field values. 
As for the case of $-$ parity,  there is no resonant behaviour, 
the averaged probabilities are practically independent of $B$, $\overline{P}^-_\eta = 2\kappa^2$, 
and thus in~this case the $n$TTM effects will be more difficult to distinguish. 
(However,  very close to the resonance $B = B'$ the probability $\overline{P}^-_\eta$ will depend 
on the direction of magnetic field also for $-$ parity case.)    
As for $+$ parity case, the~effect is substantial only when $B\sim B'$ while in 
both limits $B \ll B'$ and $B \gg B'$ we have $\overline{P}^+_\eta = 2\kappa^2$.
Thus, for~large magnetic fields $B \gg B'$ probability will be constant independent 
on the parity sign as also can be seen directly from Equation~(\ref{Ppm}).

On the other hand, assuming that $B' \neq 0$, instead of applying non-zero magnetic field, 
one can introduce some matter potential for the ordinary component. 
In this case, taking $B=0$,  the~Hamiltonians (\ref{22}) should be changed to the form
\begin{equation}\label{22prime}
{H}^\uparrow_{nn'} =
\begin{pmatrix}
2\delta & \varepsilon \pm 2\kappa\omega' \\
\varepsilon \pm 2\kappa\omega' & - 2\omega'  \\
\end{pmatrix}, \quad\quad 
{H}^\downarrow_{nn'} = \begin{pmatrix}
2\delta  & \varepsilon \mp 2\kappa\omega' \\
\varepsilon \mp 2\kappa\omega' & 2\omega'  \\
\end{pmatrix}, 
\end{equation} 
where now signs $^\uparrow$ and $^\downarrow$ indicate  the neutron spin parallel/antiparallel 
respective  to the mirror magnetic field direction ${\mathbf B}'$. 
Correspondingly the oscillation probabilities (\ref{delta}) read: 
\begin{eqnarray}\label{delta-prime}
P^{\uparrow}(t) 
= \frac{(\varepsilon \pm 2\kappa \omega')^2}{(\delta + \omega')^2+ (\varepsilon \pm 2\kappa\omega')^2 } 
\sin^2[ \sqrt{(\delta +\omega')^2 + (\varepsilon \pm 2\kappa\omega')^2} \, t] \, , \nonumber \\ 
P^\downarrow(t) =
\frac{(\varepsilon \mp  2\kappa \omega')^2}{(\delta + \omega')^2+ (\varepsilon \mp 2\kappa\omega')^2 } 
\sin^2[ \sqrt{(\delta +\omega')^2 + (\varepsilon \mp 2\kappa\omega')^2} \, t], 
\end{eqnarray} 

Thus, at~the resonance $\delta = \omega'$, the~probability $P^\downarrow$ is resonantly enhanced 
to maximal value:
\begin{equation}\label{delta-prime-av}
P^{\downarrow}(t)=  \sin^2[(\varepsilon \mp 2\kappa \omega')t] , 
\end{equation} 
while for $\delta = - \omega'$ the same occurs for $P^\uparrow$. 
Hence, by~introducing some gas with positive or negative potential 
one could achieve a resonance amplification 
of $n-n'$ conversion probability for one polarization, for~both~parities.   

Let us consider e.g.,~the case of  cold neutrons propagating in air or in another gas 
described by the positive Fermi quasi-potential $V=2\delta$ \cite{Golub:book}. 
At the same time,  the~density of this gas can be low enough  to neglect the incoherent scattering
at a finite angles and the absorption of neutrons. 
e.g., for~air at normal temperature and pressure (NTP) this Fermi potential is $V \approx 0.06$ neV. 
The probability of elastic scattering and absorption for cold neutrons in air at the NTP is
$\approx$0.05 per meter of path. It is low enough to consider few meters of 
cold neutron beam propagation in air. 
In this way, if~the polarization dependent losses will be observed, 
they can test the $n$TTM-induced $n-n'$ conversions if the mirror magnetic field $B'$ at the Earth 
is up to few Gauss.



\section{Experimental Limits  from Direct Searches with~UCN } \label{sec3}

The experiments with the UCN traps 
such as~\cite{Ban,Serebrov1,Serebrov2,Bodek,Altarev,ILL}
were performed for direct search of $n \rightarrow n'$  transition under the hypothesis of 
non-zero mass mixing $\varepsilon$.  
The above experiments study the dependence of the loss rate of the neutrons stored in the 
UCN traps  on the strength and direction of the applied magnetic field $\mathbf{B}$.  
Under the assumption of possible presence of mirror magnetic field $\mathbf{B}'$ at the Earth, 
their results can be interpreted as upper limits  on  $\varepsilon$ (or lower limits 
on $n-n'$ oscillation time $\tau_{nn'} = \varepsilon^{-1}$) as a function of mirror field $B'$. 
Below, we shall use the results of these experiments for obtaining the similar limits 
on $n-n'$ transitional moment $\eta$. 
In this section we consider more interesting case corresponding to $(+)$ parity 
leaving the more simple case of negative $(-)$ parity, 
particularly in the resonant region, for~discussion~elsewhere. 

The experimental strategy described e.g.,~in Ref.~\cite{ILL} is the following.  
In the absence of  $n-n'$ transitions, the~number of neutrons $N(t_\ast)$ surviving 
after effective storage time $t_\ast$ inside a UCN trap should not depend on $\mathbf{B}$, 
given that the usual UCN losses, such as neutron decay, wall absorption, or~up-scattering, 
are magnetic field independent according to standard physics. 
But, when we consider oscillations, during~the motion between two consecutive wall collisions 
the neutron can transform into a sterile state $n'$, and~so per each collision it has a certain probability 
to escape from the trap. 
Therefore~the number of survived neutrons in the UCN trap with applied 
magnetic field $\mathbf{B}$ after a time $t_\ast$ is given by 
$N_{\mathbf{B}}(t_\ast) =  N(t_\ast)  \exp \big[ -n_\ast  \overline{P}(\mathbf{B})\big]$,
where $N(t_\ast)$ is the amount of UCN that would have survived in the absence of $n-n'$ oscillation,
$\overline{P}(\mathbf{B})$ is the average  probability of $n-n'$ conversion 
between the wall scatterings given in Equation~(\ref{PB}) 
and $n_\ast = n(t_\ast)$ is the mean number of wall scatterings for the neutrons survived after the 
time $t_\ast$. 
The oscillation probability depends on strengths $B$ and $B'$  of magnetic fields 
and the angle $\beta$ between their directions, $\cos\beta = \mathbf{B}\cdot\mathbf{B}'/BB'$. 
If the magnetic field direction is inverted, $\mathbf{B} \to -\mathbf{B}$, i.e  $\beta \to \pi -\beta$, 
we have 
$N_{-\mathbf{B}}(t_\ast) = N(t_\ast)   \exp \big[ -n_\ast  \overline{P}(-\mathbf{B})\big]$.
Then~one can define the ``directional'' asymmetry between $N_{\mathbf{B}}(t_\ast)$ and 
$N_{-\mathbf{B}}(t_\ast)$ as
\begin{equation}
A_{\mathbf{B}}(t_\ast) = \frac{N_{-\mathbf{B}}(t_\ast) -
	N_{\mathbf{B}}(t_\ast)} {N_{-\mathbf{B}}(t_\ast)+N_{\mathbf{B}}(t_\ast)} = 
\frac{n_\ast}{2}
\big[ \overline{P}(\mathbf{B}) - \overline{P}(-\mathbf{B}) \big]  
= n_\ast \overline{\cal D}(B)  \cos\!\beta ,
\label{eq:AB}
\end{equation}
where $\overline{\cal D}(B) \cos\beta$  
is the difference between the respective average oscillation probabilities  which is 
proportional to $\cos\beta$ and the expression of $\overline{\cal D}(B)$  is given in Equation~(\ref{Final}). 
The common factor $N(t_\ast)$ cancels in the neutron count ratios, and~$A_{\mathbf{B}}$ 
directly traces the difference of the average oscillation probabilities 
in magnetic fields of the opposite directions, $\mathbf{B}$ and $-\mathbf{B}$.

One can also compare the average between those two counts:
$N_B(t_\ast) = \frac12 \big[N_{\mathbf{B}}(t_\ast) + N_{-\mathbf{B}}(t_\ast)\big]$ 
with the counts $N_0(t_\ast)$ acquired under zero magnetic field. 
We then obtain
\begin{equation}
E_{B}(t_\ast) = \frac{N_0(t_\ast) - N_{B}(t_\ast)}
{N_{0}(t_\ast)+N_B(t_\ast)} = n_\ast \left[\frac12 \overline{P}(\mathbf{B}) +\frac12 \overline{P}(-\mathbf{B})
- \overline{P}(0)\right] 
=   n_\ast \big[ \overline{\mathcal{P}}(B) - \overline{\cal P}(0) \big] , \label{eq:EB}
\end{equation}
where $\overline{\mathcal{P}}(B)$
is the average of the oscillation probabilities between two opposite directions, see~Equation~(\ref{Final}). 
Thus, the~value $E_{B}$ measuring the difference between the average probabilities 
at zero and non-zero magnetic fields should not depend on the magnetic field orientation 
but  only on its modulus $B = \vert \mathbf{B} \vert$.

\begin{figure}[H]
\centering
	\includegraphics[width=7.5cm]{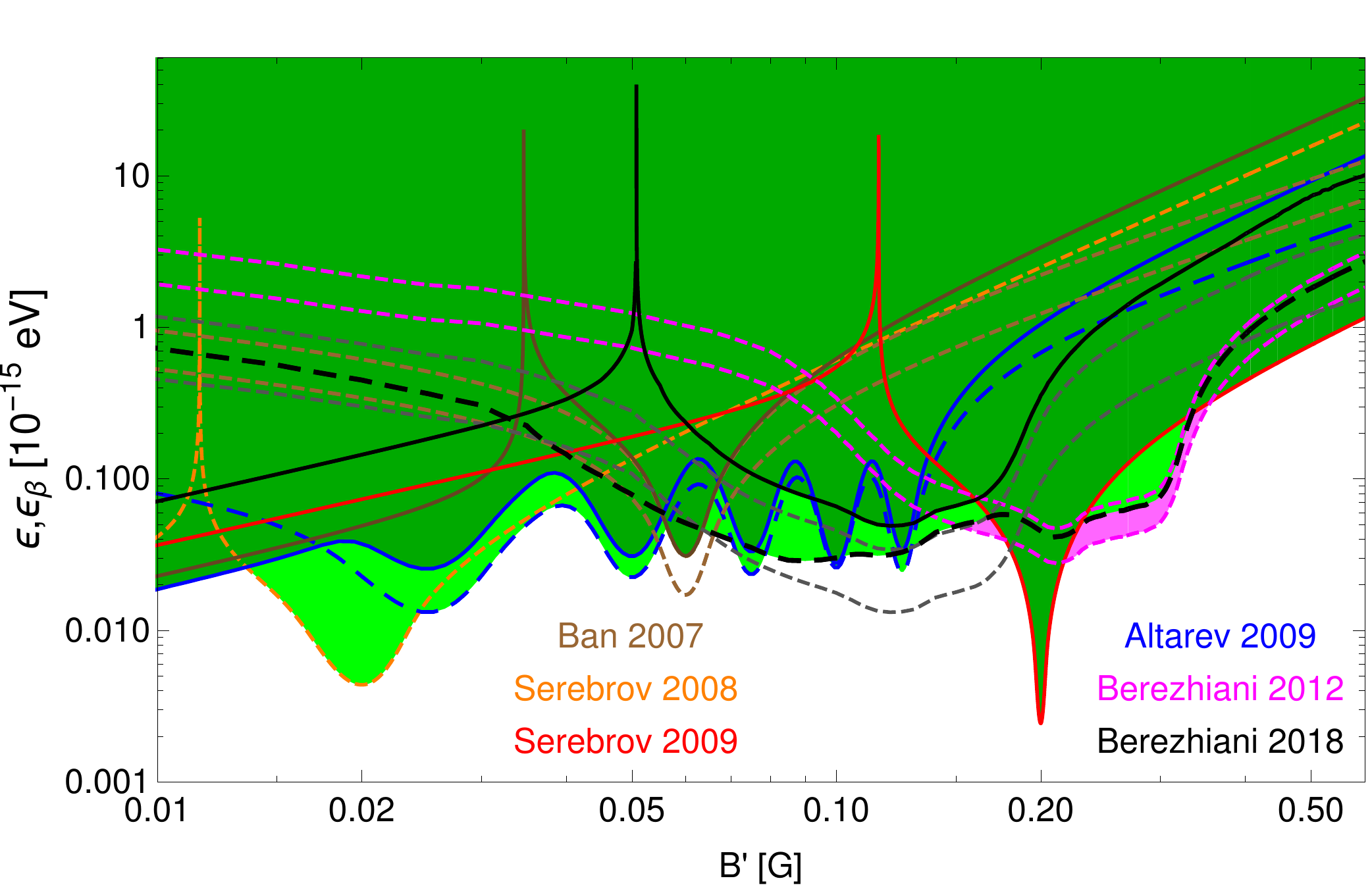}
	\includegraphics[width=7.5cm]{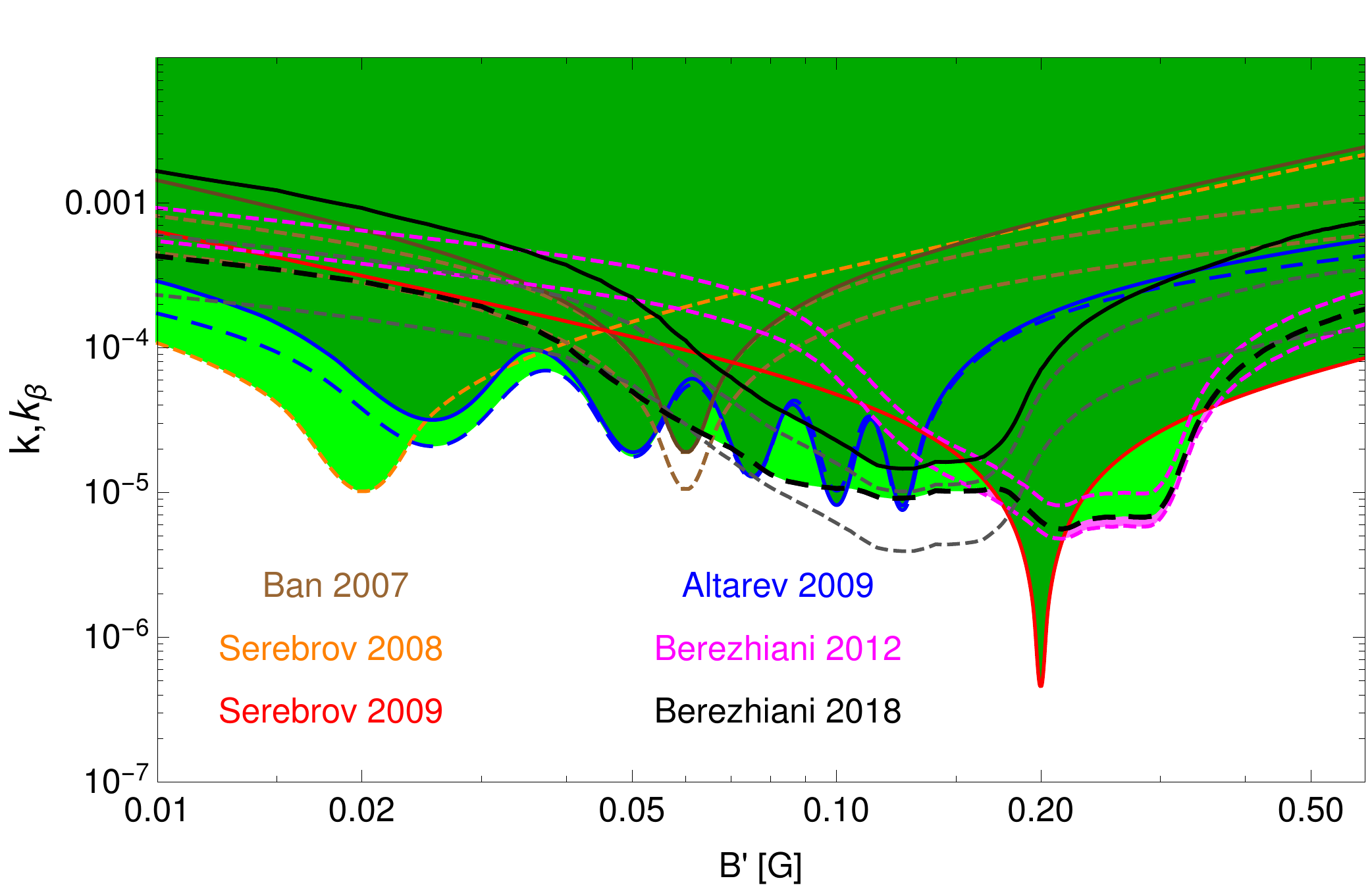}
	\caption{\label{fig:limits}
		\small  Upper limits for the oscillation parameters $\varepsilon$, $\varepsilon_\beta$ 
		assuming $\kappa = 0$ (\textbf{left panel}) and
		$\kappa$, $\kappa_\beta$~assuming $\varepsilon = 0$ (\textbf{right panel}) 
		as a function of the mirror magnetic field $B'$.
		The parameter areas  excluded at 95 \% C.L.  by the experiments
~\cite{Ban,Serebrov1,Serebrov2,Bodek,Altarev,ILL}  for $\varepsilon$($\kappa$)  
		are confined by solid curves of respective colors 
		and  are shaded in dark green, and~for $\varepsilon_\beta$($\kappa_\beta$) 
		are confined by dashed curves of respective colors and shaded  by light green.  
		Dashed pink curves confine the parameter area which can be relevant for $5.2\sigma$ anomaly  in $A_B$ 
		reported in Ref.~\cite{Nesti}, and~its part still not excluded by other data is shaded in pink.  
		Let us remark, that the exclusion areas 
are obtained by assuming that the magnitude of mirror field $\mathbf{B}'$ 
and/or its direction with respect to the laboratory site remained unchanged during several years 
of  time lapse between  experiments~\cite{Ban,Serebrov1,Serebrov2,Bodek,Altarev,ILL} 
which however might not be the case. 
}
\end{figure}

The current experimental situation in the case of $n-n'$ mass mixing $\varepsilon$ 
is summarised in Ref.~\cite{ILL}. These experiments,  by~measuring $E_B$ for different values 
of $B$, in~fact determine the mass mixing parameter $\varepsilon$ 
while via measuring $A_{\mathbf{B}}$ they determine the combination 
$\varepsilon_\beta = \varepsilon \sqrt{\vert \cos\beta \vert }$, 
i.e., the mass mixing corrected by the unknown angle $\beta$ between
the ordinary magnetic field $\mathbf{B}$ and background mirror field $\mathbf{B}'$. 
Namely, the~results of all dedicated experiments~\cite{Ban,Serebrov1,Serebrov2,Bodek,Altarev,ILL} 
were used to set lower limits on the oscillation time $\tau=\varepsilon^{-1}$
and the combined value $\tau_\beta = \tau/\sqrt{\vert \cos\beta \vert}$ as a function of mirror 
magnetic field $B'$, see  Figure~7 of Ref.~\cite{ILL}  (in the limit $B'=0$, the~
upper limit corresponds to $\tau > 448$ s at 90 \% C.L.~\cite{Serebrov1}).  
For convenience, in~this paper (see left  panel of Figure~\ref{fig:limits})  we show the same 
limits directly in terms of physical parameters $\varepsilon$ and $\varepsilon_\beta$ as upper limits. 
(Clearly, in~our generic context, these~limits correspond to the case without $n$TTM 
contribution, i.e.,~$\kappa=0$.) 

The data of the same measurements can be used for setting the upper limits on 
$\eta = \kappa \mu$ assuming in turn that $n-n'$ mass mixing is vanishing and 
$n-n'$ conversion occurs solely due to  $n$TTM effects. In~the right panel of Figure~\ref{fig:limits} 
we show upper limits on $\kappa$ and $\kappa_\beta = \kappa \sqrt{\vert \cos\beta \vert}$ 
in the case of $\varepsilon = 0$.  
We used the same Monte Carlo code developed in~\cite{Biondi:2018fmr} and used in~\cite{ILL} 
to calculate the average oscillation probability $\overline{P}_{BB'}$ for the UCN inside the trap in 
non-homogeneous magnetic field but using the probabilities $P_\eta$ 
instead of $P_\varepsilon$. 
This will give us differences between the shapes of the exclusion regions for mass mixing 
and $n$TTM cases which can be observed by comparing the left and right panels of  Figure~\ref{fig:limits}.
The magenta shaded areas in Figure~\ref{fig:limits} correspond to parameter regions relevant 
for more than $5\sigma$ deviation from null hypothesis in the measured  asymmetry 
$A_B =(6.96\pm 1.34)\times 10^{-4}$~\cite{Nesti} 
which still is not excluded  by the present experimental limits. 
Let us remind that this anomaly, as the variation of  between the numbers of the survived UCN 
stored under the vertical magnetic field $B\approx 0.2$~G directed  up and down, 
was obtained in Ref.~\cite{Nesti}  via re-analysis  of the experimental data of Ref.~\cite{Serebrov2}.  
This asymmetry can been interpreted as an effect of $n\rightarrow n'$ oscillation in the presence 
of a mirror magnetic field on the order of 0.1 G at the Earth \cite{Nesti,ILL}.


In this section we have not considered the effects of magnetic field gradients 
to be introduced in the following Section~\ref{sec4} for the estimates of possible magnitude of 
$n$TMM. In~UCN experiments~\cite{Ban,Serebrov1,Serebrov2,Bodek,Altarev,ILL}
certain attempts were made to maintain the uniformity of magnetic fields,
so the effects of collapse of $(n,n')$ system in the trap volume between the wall 
collisions due to gradients might be not very significant besides the fact that detailed 
field maps for most of these experiments are also not available. Thus, the~limits shown 
on the right panel in  Figure~\ref{fig:limits} should be considered as orientative 
upper limits for $\kappa$ obtained in the assumption of the presence of mirror magnetic field $B'$.


\section{\boldmath{$(n,n')$} System in Non-uniform Magnetic Field}\label{sec4}

If at $t \leq 0$ the $(n,n')$ system were in vacuum and in a constant magnetic field $B = B_0$ ($B'=0$), 
then~for initial condition at $t=0$, the~kinetic energies $T=T'$ are the same, 
as~it is assumed in Equation~(\ref{H1}). 
The~system will oscillate in time between two states $n$ and $n'$ 
which  interact with ordinary matter differently.
Two orthogonal eigenstates of the oscillating ($n,n'$) system are formed corresponding 
to the different total energy eigenvalues. Namely, the Hamiltonian 
eigenstate $n_1 = \cos\theta \, n + \sin\theta \, n'$ 
has interaction properties close to that of the ordinary neutron ($n$) 
and the state $n_2 = \cos\theta \, n' - \sin\theta \, n$ close to that of the mirror neutron ($n'$)  
once the mixing angle $\theta$ arising due to non-diagonal terms in Hamiltonian 
is small. If the medium is uniform and also constant in time, the eigenstates $n_1$ and $n_2$ 
propagate independently. 
Let us remind however that in non-uniform medium the angle $\theta$ 
is  position dependent and thus the notion of the Hamiltonian eigenstates 
has only a local significance. However if the evolution of the system is adiabatic, 
then the Hamiltonian eigenstates adiabatically evolve from one position to another. 

Let us assume that at $t > 0$  the system enters a region of non-uniform magnetic field 
which is a function of coordinates and has non-zero gradients. 
The gradient of potential energy  $U=(\boldsymbol{\mu} \cdot \bf{ B})$ 
from a semi-classical point of view generates a force. 
In some UCN experiments, e.g.,~in the UCN$\tau$ experiment with 
magnetic Hallbach array~\cite{Pattie:2017vsj}, this force can exceed the Earth gravitational attraction 
force and cause vertical bouncing of neutrons (of a certain polarization) from the surface 
with an arranged strong magnetic field gradient. 
By virtue of the Mirror Model this gradient force mainly acts on the ``almost" neutron component 
$n_1\simeq n$ of the system but very weakly on the ``almost" mirror component $n_2 \simeq n'$, 
while the gravity is the same for both components $n_1$ and $n_2$. 
A similar situation occurs in the collision of neutrons with the wall in gravitational UCN traps, 
where the gradient is due to the Fermi potential of the trap walls. 
A positive Fermi potential $V_F$ of the wall material will repulse neutrons with kinetic energy 
smaller than $V_F$ at the characteristic distance of the neutron wave packet, 
$\sim$$1000~\mbox{\normalfont\AA}$  for typical $V_F=100$ neV. 
In this way, the~gradient of potential in the interaction with wall can be estimated as $\sim$$1$ eV/m. 
In the UCN$\tau$ experiment~\cite{Pattie:2017vsj} the gradient of the magnetic field is $\sim$1 T/cm near 
the bouncing surface that corresponds to the gradient of potential energy $\sim$$6 \times 10^{-6}$ eV/m.

Integrated over time $\Delta t$, this force provides a relative momentum $\Delta p$ between the two components. 
If it exceeds the width of the neutron wave packet $\Delta p_{p}$, then that will lead to the
``separation of components'' or to decoherent collapse of $(n,n')$ system 
into either $n $ or $n'$ states. 
If this force would be exerted only for time $\Delta t$ and then reduced to zero, then finite relative momentum  
$\Delta p > \Delta p_{p}$ will continue in time the separation of the components. 
The same conclusion can be formulated in a language more appropriate for quantum mechanics 
as a potential energy variation different for the two components 
of $(n,n')$ wave function in non-uniform magnetic~field.

At some point the two components will be ``pulled away'' by gradient, and~the entanglement of $(n,n')$ system 
moving in the non-uniform magnetic field will be broken. 
If~``measured'' at this moment the $(n,n')$ system will be found with probability $P$
 in the pure state of  ``mirror neutron'' $(0,1)$ or with probability $1-P$ 
 in the state of ``neutron'' $(1,0)$. If~not ``measured,'' 
 each of the states will be subject to further oscillation 
 with the reset of initial conditions at the time of ``separation.''

The situation is similar in UCN traps  with the material walls 
where entanglement will be broken with $n$ component being reflected or absorbed 
by the wall with the probability $1-P$ and $n'$ component escaping from the UCN trap 
through the~wall with a small probability  $P$.

More generally, the~motion of the oscillating $(n,n')$ system in the slowly (adiabatically) changing 
magnetic field can lead either to a change of the kinetic energy of the components of wavefunction 
and/or to their spin rotation. 
Both these effects can result in decoherence of the system. The~decoherence problem expectedly 
can be properly treated by modern methods using the density matrix formalism~\cite{Lindblad1,Lindblad2}. 
One can consider master equation of the evolution of density matrix of $(n,n')$ system 
in the environment of the external potential $U(\bf{r})$, different for two components of the system
(four components when spins are included). 
These techniques  are not yet sufficiently developed and tested in respect to particle oscillations 
and therefore we will refrain from the construction and solution of the density matrix evolution equation.
Instead we try to consider the decoherence problem qualitatively. 
For~these qualitative arguments we will ignore the spin rotation in a non-uniform magnetic field 
and will consider one-dimensional motion in a field with a gradient. 
Let us first assume a hypothetical simple case of a constant gradient of the magnetic field along 
the direction of the motion. 
The magnitude of $B$ is linearly increasing with the distance $x$, but~the direction of vector $\bf{B}$ 
remains practically the same. For~a neutron with a certain polarization propagating along axis  $x$
from the initial condition $(n,n') = (1,0)$ and for the observation time $\Delta t$, 
the kinetic energy width of the wave packet $\Delta E_p$ of the $(n,n')$ system is
\begin{equation}
\label{eq:9}
\Delta E_p \sim 1/{\Delta t} .
\end{equation}

For a path $\Delta x = x_f - x_i$, passed by the neutron for the observation time $\Delta t$, 
the change of potential energy produced by the magnetic field gradient is  $\Delta U = U_f - U_i$. 
If $\Delta U$ is  much smaller than $\Delta E_p$,
\begin{equation}
\label{eq:10}
\Delta U = \mu \Delta B \ll \Delta E_p , 
\end{equation} 
then the $(n,n')$ system will remain~entangled. 

For sufficiently large $\Delta B$, the~entanglement of the $(n,n')$ system will be broken. 
This means that the system will collapse to pure states of either $n'$ or $n$ with probabilities 
$P$ and $1- P$ correspondingly. Assuming that the velocity will not change 
significantly by magnetic gradient for the time $\Delta t$ (say, for~the time of flight between
 two wall collisions in a UCN gravitational trap), then the gradient that should lead to the decoherence collapse 
 of the system can be estimated in a following way:
\begin{equation}
\label{grads}
\frac{\Delta B}{\Delta x} > \frac{1}{\mu v (\Delta t)^2} = \frac{v}{\mu (\Delta x)^2} .
\end{equation} 

Thus, for~the entangled evolution of the $(n,n')$ system e.g.,~with velocity $v =3$ m/s
in a UCN gravitational trap with a flight time $\Delta t \simeq 0.1$ s between wall collisions, 
the magnetic field gradient should be $\ll 3.7$ mG/m. 
Larger gradients can cause the collapse of the wavefunction at an earlier time 
and can cause the transformation of $n$ to $n'$ to occur in the volume of the trap 
rather than in collisions with the trap walls. 
Thus, the~effect of a non-uniform magnetic field might result in the $n \rightarrow n'$ 
transformation in the trap volume with the same result as that which would be expected 
in collisions of the entangled $(n,n')$ system with the trap walls. 

In the UCN gravitational trap experiments measuring the neutron lifetime 
the Earth magnetic field $B \simeq 0.5$~G is usually considered as a non-essential factor. 
The number of wall collisions is experimentally extrapolated to the ``zero number of collisions,''
e.g., in~\cite{Serebrov:2004zf,Serebrov:2017bzo,Serebrov:2007ve}. 
A magnetic field non-uniformity can produce a similar disappearance effect 
in the volume of UCN gravitational traps, but~this effect is not removable by extrapolation 
to the zero number of wall collisions. 
Unfortunately, the~UCN gravitational trap experiments do not use magnetic shielding of the trap, 
and the actual maps of magnetic fields in these experiments are unknown. 
For getting some idea of the possible non-uniformity of the Earth magnetic field in such conditions, 
we have measured some vertical gradients as high as 75 mG/m in an arbitrary general-purpose 
laboratory room of a typical university~building. 
Therefore, we can advocate that
the contribution of $n$TMM (that outside the resonance does not depend on the magnitude 
of the magnetic field) to $n \rightarrow n'$ transformation could be also essential 
as a source of the neutron~loses. 


In the proposed cold-neutron-beam disappearance/regeneration 
experiments~\cite{Berezhiani:2017azg, Broussard:2017yev}, with~average neutron velocity 
$v \sim 800$ m/s and  flight path e.g.,~$16$ m, for the entangled evolution of the $(n,n')$ oscillating system  
the gradients $\ll$$0.4$ mG/m would be required. 
Larger gradients can shorten the path of entangled evolution and 
effectively will lead to multiple shorter-in-time collapses which will increase the production of $n'$ states, 
since the probability in the fields larger than few mG and far from
resonance due to $n$TMM will remain constant $2\kappa^2$.



\section{Possible Magnitude of Neutron~TMM} 

It is a well known problem in UCN gravitational trap experiments that the measured wall losses 
are larger than these predicted from theoretical models using known scattering lengths of materials 
(see discussion in Ref.~\cite{Pokotilovski-2016} and references therein). 
Lowering the temperature of the trap walls, although~reducing the loss coefficient (per single collision), 
does not resolve the discrepancy between experiment and theoretical calculations. 
Only in one of a few experiments using fomblin-oil coated trap~\cite{Serebrov:2007ve} 
the measured ($\approx 2 \times 10^{-6}$) and expected ($\approx1\times 10^{-6}$) loss factors 
per wall collision were in good agreement. (In all other experiments the measured losses 
were significantly exceeding the theoretical predictions.) 
Hence, one can speculate that losses at least at the level of $1 \times 10^{-6}$ per collision can be due 
to the losses in the trap volume caused by the neutron TMM 
in the non-uniform environmental magnetic field. 
The local Earth magnetic field in these experiments can be affected by magnetic constructional materials, 
platforms, etc., as~well as by metal reinforced concrete walls of the industrial buildings. 
Let us assume that in a typical trap of the size $\sim$$0.3$ m the magnetic field uniformity is $<$$5\%$ 
from wall to wall, such that the moving neutron can see a constant gradient of about 75 mG/m. 
We are not considering the effect of neutron spin rotation due to possible change of the direction 
of the magnetic field---this might result in additional decoherence effects that are not discussed here. 
For a typical UCN velocity 3 m/s, and~assuming that the velocity  will not essentially be changed 
by the gradient during the flight from wall to wall, from~Equation~(\ref{grads}) we can find that 
the typical time for collapse to occur will be $\sim$$0.02$ s.  
Thus, we can estimate that $\sim$$5$ volume collapse events will occur per one collision with the trap wall. 
Attributing this totally to the neutron TMM transformations with constant probability $2\kappa^2$, 
we can estimate that losses per collision of $1 \times 10^{-6}$ correspond to a neutron TMM of
\begin{equation} \label{etawall}
\kappa \simeq 3 \times 10^{-4}
\end{equation}

In the neutron lifetime measurement with large UCN gravitational trap~\cite{Serebrov:2017bzo}, 
assuming the same magnetic field gradients, the~number of decoherence events per second occurring 
in the volume of the storage trap can be estimated as $\sim$$50$ s$^{-1}$. With~the probability of 
$n \rightarrow n'$ transformation per event as $2\kappa^2$, this can generate a neutron disappearance 
rate that can explain the difference~\cite{Pokotilovski:2006gq} between the result~\cite{Serebrov:2017bzo} 
of the UCN disappearance lifetime experiment and the result of the beam appearance measurement~\cite{Yue:2013qrc}. The~required value of $n$TMM in this case can be estimated again as (\ref{etawall}).

If magnetic field gradients in the gravitational trap are higher than we have assumed, 
then the magnitude of neutron TMM required to produce mentioned disappearance rate can be 
lower than in (\ref{etawall}). 
Also, if~$|B'|<|B|$ is present it might increase the conversion probability thus reducing
our estimate of $\kappa$  in (\ref{etawall}). Would the UCN lifetime experiment with gravitational trap 
be magnetically shielded with residual field magnitude $\ll 1$ mG 
(that means assuming that both $B$ and $B'$ are vanishing) 
then $n \rightarrow n'$ effect due to $n$TMM will vanish 
and measured lifetime might be affected only by a smaller effect of $n \rightarrow n'$ oscillations due to 
mixing mass $\varepsilon$ as was discussed~earlier. 

In experiments with ``UCN magnetic field trap'' \cite{Pattie:2017vsj, Ezhov:2014tna}, where neutrons are repulsed
from the strong gradient of the magnetic field, the~number of decoherence events can be greatly increased. 
For~a rough estimate we have attempted to reproduce in a simplified one-dimensional way 
the vertical bouncing of UCN in the magnetic field with a strong gradient described 
in the papers~\cite{Pattie:2017vsj} up to the height of 50 cm and obtained with Equation~(\ref{grads}) 
approximately 3500 decoherence events per a second of the neutron motion. 
Thus, to~produce a UCN disappearance rate $\sim$$10^{-5}$ per second in experiment~\cite{Pattie:2017vsj}, 
using constant conversion probability $2\kappa^2$ one can estimate the magnitude of  $n$TMM:
\begin{equation} \label{ucntau}
\kappa \approx 4.4 \times 10^{-5}
\end{equation}

Simulations for neutron propagation in the trap with more details of magnetic field configuration 
(not available to us) will likely affect this estimate. 
Also, the~potential reason for disappearance in~\cite{Pattie:2017vsj} can be a change 
of the spin precession phase due to the presence of $(n,n')$ oscillations~\cite{Berezhiani:2008bc} 
that can lead to unexpected depolarization effect in the magnetic trap. 

We also can get another $n$TMM estimate from the different interpretation of the limit on $\varepsilon$ 
obtained in~\cite{Serebrov2} at $B=0$ and under assumption that $B'=0$. 
We will assume for this result instead that mirror magnetic field $B'$ is present (although unknown) 
with magnitude larger than few mG and thus the measured limit on disappearance probability 
in~\cite{Serebrov2} should be taken as a limit determined by the magnitude of probability $2\kappa^2$ 
due to $n$TMM. From~this we can obtain the following estimate:
\begin{equation} \label{slimit}
\kappa < 2 \times 10^{-4}
\end{equation}

We can conclude this section by noting that these estimates, although~very rough, are 
suggesting the order of magnitude of the $n$TMM 
$\kappa \sim 10^{-4} - 10^{-5}$ that doesn't contradict 
the existing experimental observations and might be the parameter 
of the mechanism responsible for the neutron disappearance in 
the UCN trap lifetime experiments, such as~\cite{Pattie:2017vsj,
	Serebrov:2017bzo, Ezhov:2014tna} and others. Such range of possible 
magnitudes for $\kappa$ is also allowed by the limits obtained from the 
direct $n\rightarrow n'$ search discussed in Section~\ref{sec3}.


\section{How nTMM can be~Measured} 

Beyond the results obtained in $n \rightarrow n'$ direct searches with UCN~\cite{Ban,Serebrov1,Serebrov2,Bodek,Altarev,ILL,Nesti} and summarized in~\cite{ILL} the
possible new searches with cold neutrons, described in~\cite{Berezhiani:2017azg,Broussard:2017yev}, 
might bring more evidence whether $n \rightarrow n'$ transformation exists. 
These new proposed measurements are based on the detection of cold neutrons
 in intense beam after coming through the ``region A'' of controlled uniform magnetic field 
 (in~disappearance mode) or on total absorption of the cold neutron beam after 
passing through the ``region A'', allowing only produced mirror neutron to pass
through an absorber and then regenerating them back to the detectable neutrons 
in the second ``region B'' with similar controlled uniform magnetic field (regeneration mode). 
Variation of the magnitude of uniform magnetic field $B$ in the range 0 to 0.5 G by small-steps scan 
(also with possible variation of direction of vector $\bf{B}$) can reveal the resonance 
in the $n$-counting rates corresponding to $|B| \eqsim |B'|$ with magnitude and width 
related to mass mixing parameter $\varepsilon$ in Equation~(\ref{H2}). 
The~presence of $n$TMM can enhance the resonance magnitude or, 
as~follows from Equation~(\ref{Final}), can offer an alternative interpretation 
of the effect (if observed or excluded) in terms of magnitude of $n$TMM.

We would like to discuss here other two new methods that can be used for 
the detection of $n$TMM effect in the magnetic fields stronger than the Earth 
magnetic~field. 

The first method is a variation of the regeneration 
method where ``region A'' and ``region B'' mentioned above, instead of 
uniform magnetic field, are implementing strongly non-uniform magnetic 
field. Since $n$TMM transition probability remains constant 
in any sufficiently large magnetic field, strong gradients of the field can 
generate large number of decoherent collapses of $(n,n')$ system into 
either $n$, or~with a small probability to $n'$-states; the latter
will lead to enreachment of $n'$ in the ``region A.'' The absorber between
``region A'' and ``region B'' can remove all not-transformed neutrons from 
the beam allowing only the $n'$ to pass through the absorber. Strong gradients 
of the magnetic field in the ``region B'' will similarly enhance the transformation 
of $n'$ back to detectable $n$, the~latter can be counted above the natural 
background e.g.,~by the $^{3}$He~detector.

Strong magnetic field gradients in ``region A'' and ``region B'' can be implemented e.g.,~as 
a series of equally spaced coils around the neutron beam with alternating directions of constant current. 
Every coil will contribute opposite direction of magnetic field with zig-zag pattern and can provide 
practically constant strong $|dB/dz|$ gradient along the beam axis. Simple estimates show 
that the gradient $\sim$$100$ G/m can be maintained along the vacuum tube that is e.g.,~15 m long. 
For~the cold neutron beam with average velocity $\sim$$800$ m/s using Equation~(\ref{grads}) 
we can estimate that the number of decoherent collapses in such ``region A''- or ``region B''- devices 
will be around~500. Thus, the number of regenerated neutrons can be enhanced by factor $(500)^2$.

The use of superconducting solenoidal magnet with a borehole along the beam axis and 
with maximum magnetic field of $\sim$$10$ T will be a more compact approach for reaching high gradients. 
The~total length of the magnet can be of the order of 1 m. 
The~magnet front field ramping-up side can serve as a ``region A'' and back side with ramping-down 
field as a ``region B,'' each with a decoherence collapse factor $\sim$$700$. 
A~beam absorber should be installed in the middle of the magnet, thus transforming it into a  regeneration~device.

With cold neutron beams available e.g.,~at ILL and at HFIR/ORNL reactors, or~at SNS/ORNL, 
or~at the future ESS spallation neutron source~\cite{nsources}, where cold beam intensities are in the range $10^{10}$--$10^{11}$ n/s, the~effect of $n$TMM
in the whole range of $\kappa \sim 10^{-4}$--$10^{-5}$ 
can be explored in rather short and not expensive experiments~\cite{nordita}.

The second method is based on the idea of compensation of Fermi quasi-potential of the gas media 
with a constant uniform magnetic field as discussed in the Section~\ref{sec2} of this paper with oscillation 
probability given by Equations~(\ref{delta}) and (\ref{delta-prime-av}). In~Equation~(\ref{delta}) both parameters 
of the resonance $\delta$ and $\omega$ are known and can be adjusted to compensate each other.
Then, close to the resonance the probability of quasi-free oscillation can coherently grow as square of 
neutrons propagation time through the gas, e.g.,~through the air at NTP. This will work, according to 
Equation~(\ref{delta}), only for one of neutron polarizations. Regeneration scheme can be used here again with 
``region A'' and ``region B'' represented by the air-filled tubes of length e.g.,~$L=2$ m in the solenoidal 
constant uniform field of $B$$\sim$$10$~G. Effects of the beam scattering and absorption in the gas will 
reduce the rate of regenerated neutrons by $\approx$$20 \%$. Probability of regeneration observation 
provided by this second method can be an order of magnitude higher than for the first method discussed 
above. If~the resonance will be observed for the values $(\delta - \omega)$ different from anticipated zero, 
that will indicate the presence of either mirror magnetic field $B' < B$ or small mirror Fermi-potential $\delta'$. 
In this case, the~same two-tubes layout can be used for regeneration search with the magnetic field $B$ 
shielded below a $\sim$mG and the gas pressure in the tubes, i.e.,~parameter $\delta$ varied in order to 
find the resonance condition described by Equation~(\ref{delta-prime-av}) corresponding to the unknown mirror 
magnetic field $B'$. The~second method can be further optimized for the use with a beam of collimated 
UCN propagating in the gas with low absorption. In~this case, due to slow UCN velocities the probability for 
$n \rightarrow n'$ transformation potentially can grow approaching the maximum~value. 


\section{Conclusions} 

Generally speaking, in this paper we discussed the possible environmental effects
on the oscillating quantum system which environment can be related to dark matter. 
The paradigm of  the neutron--mirror neutron transition~\cite{Berezhiani:2005hv}
 should naturally comprise additional effects that are inherent for mirror matter which 
 essentially is a self-interacting atomic dark matter.
This sort of dark matter, composed dominantly by lighter mirror nuclei as hydrogen and helium, 
is not easily accessible in direct dark matter search  (see however Refs.~\cite{DAMA}).
But its collective environmental effects could have an interesting consequences for the fundamental 
particle properties and influence the values of the physical quantities  such as particle lifetimes, 
spin precession frequencies, rates of baryon violating processes, etc. which effects are not yet 
adequately explored in the high precision particle physics experiments.

For the neutron--mirror neutron transitions which can be caused by the mass mixing and/or by
transition magnetic (or electric) dipole moment between $n$ and $n'$ states,   these
environmental effects can be related to the~existence of mirror magnetic fields on Earth~\cite{Berezhiani:2008bc},
the~possible accumulation of mirror matter in the solar system and inside the Earth
which could also form the mirror atmosphere of the latter~\cite{AntiDM},
and~possibly other more exotic effects as long range fifth-forces \cite{BL} 
or gravity forces in bigravity theories \cite{bigrav},  
which can differently act on ordinary and mirror matter components. 
In difference from the ordinary matter environment which can be controlled
(the gas can be pumped out for reaching a good vacuum and the magnetic field can be screened),
environment of dark matter cannot be altered  in experiments. 
Nevertheless, these effects can be probed experimentally.

The~existing difference between the neutron lifetime measurements 
by the beam (appearance) experiments \cite{Yue:2013qrc}
and the UCN trap (disappearance) 
experiments~\cite{Pattie:2017vsj, Serebrov:2004zf, Serebrov:2017bzo, Serebrov:2007ve, Ezhov:2014tna}
can be explained by $n - n'$ conversions due to small  transition magnetic moment.
The~magnitudes of $n$TMM extracted from the difference in neutron lifetime results 
can be measured in rather simple experiments,
such as one proposed in~\cite{Broussard:2017yev} for the GP-SANS beamline at HFIR.
If~the mirror magnetic field $B'$ with non-zero value is  present,
the~experimental strategy with controlled uniform magnetic field $B$ can be pursued,
as  described in the papers~\cite{Berezhiani:2017azg,Broussard:2017yev}. 
This would allow the detection of the $n\rightarrow n'$ process and could also reveal the environmental
effects of mirror matter e.g. by determining the magnitude and the direction of mirror magnetic field.
This might be a venue for observation of new effects which are usually ignored in experiments.   

\vspace{6pt} 

\section{Acknowledgements}

\noindent 
We would like to thank Josh Barrow, Leah Broussard, Sergey Ovchinnikov, 
Yuri Poko\-tilovski,  Anatoly Serebrov, George Siopsis and Arkady Vainshtein for useful discussions. 
		Z.B. and Y.K, acknowledge the hospitality provided by the Institute for Nuclear Theory, University of Washington, 
	Seattle, where in October 2017 this work was initiated during the Workshop 
	INT-17-69W  ``Neutron Oscillations: Appearance, Disappearance, and~Baryogenesis.'' 
	The work of Z.B. was supported in part by the triennial research grant No. 2017X7X85K 
	``The Dark Universe: Synergic Multimessenger Approach'' 
	under the program PRIN 2017 funded by the Ministero dell'Istruzione, Universit\`a e della Ricerca (MIUR)
	of Italy,  
	and in part by  triennial research grant DI-18-335/New Theoretical Models for Dark Matter Exploration 
	funded by Shota Rustaveli National Science Foundation (SRNSF) of Georgia. 
	The work of Y.K. was partially supported by US DOE Grant DE-SC0014558.  
	The work of L.V. was supported by the NSF Graduate Research Fellowship under Grant No.~DGE-1746045.

\end{document}